\documentclass{vldb}

\usepackage{graphicx}
\usepackage{balance}  

\usepackage{times} 
\usepackage{tipa}

\usepackage{hyperref}
\usepackage{color}
\usepackage{listings}

\definecolor{light-gray}{gray}{0.98}
\definecolor{dark-gray}{gray}{0.5}
\lstMakeShortInline^

\def\aroundlstskip{0ex}
\lstdefinestyle{noskip}{aboveskip=\aroundlstskip,belowskip=\aroundlstskip,abovecaptionskip=\aroundlstskip,belowcaptionskip=\aroundlstskip}
\lstnewenvironment{code}{}{}

\newcounter{ddlcnt}
\lstnewenvironment{ddl}[2][]{
  
  \setcounter{lstlisting}{\value{ddlcnt}}
  \lstset{style=noskip,captionpos=b,float=h!t,caption=#1,label=#2}
}{
  \addtocounter{ddlcnt}{1}
}

\newcounter{querycnt}
\lstnewenvironment{query}[2][]{
  
  \setcounter{lstlisting}{\value{querycnt}}
  \lstset{style=noskip,captionpos=b,float=h!t,caption=#1,label=#2}
}{
  \addtocounter{querycnt}{1}
}

\newcounter{updatecnt}
\lstnewenvironment{update}[2][]{
  
  \setcounter{lstlisting}{\value{updatecnt}}
  \lstset{style=noskip,captionpos=b,float=h!t,caption=#1,label=#2}
}{
  \addtocounter{updatecnt}{1}
}

\newcommand{\flwor}{\textsl{\textbackslash'fla$\dot{\mbox{u}}$(-\textreve)r\textbackslash}}

\newcommand{\eat}[1] {}

\def\betweenfigureandcaption{-1ex}

\title{AsterixDB: A Scalable, Open Source BDMS
\vspace*{-2ex}}

\newcommand{\uci}{\textsuperscript{1}}
\newcommand{\cloudera}{\textsuperscript{2}}
\newcommand{\google}{\textsuperscript{3}}
\newcommand{\ibm}{\textsuperscript{4}}
\newcommand{\marklogic}{\textsuperscript{5}}
\newcommand{\pivotal}{\textsuperscript{6}}
\newcommand{\ucr}{\textsuperscript{7}}
\newcommand{\hp}{\textsuperscript{8}}
\newcommand{\oracle}{\textsuperscript{9}}
\newcommand{\prevuci}{\textsuperscript{*}}
\newcommand{\prevucr}{\textsuperscript{\S}}

\newcommand{\auth}[2]{\mbox{#1#2}}

\author{
\alignauthor 
\auth{Sattam Alsubaiee}{\uci}
\auth{Yasser Altowim}{\uci}
\auth{Hotham Altwaijry}{\uci}
\auth{Alexander Behm}{\cloudera\prevuci}
\\
\auth{Vinayak Borkar}{\uci} 
\auth{Yingyi Bu}{\uci}
\auth{Michael Carey}{\uci}
\auth{Inci Cetindil}{\uci}
\auth{Madhusudan Cheelangi}{\google\prevuci}
\\
\auth{Khurram Faraaz}{\ibm\prevuci}
\auth{Eugenia Gabrielova}{\uci}
\auth{Raman Grover}{\uci}
\auth{Zachary Heilbron}{\uci}
\\
\auth{Young-Seok Kim}{\uci}
\auth{Chen Li}{\uci}
\auth{Guangqiang Li}{\marklogic\prevuci}
\auth{Ji Mahn Ok}{\uci}
\auth{Nicola Onose}{\pivotal\prevuci}
\\
\auth{Pouria Pirzadeh}{\uci}
\auth{Vassilis Tsotras}{\ucr}
\auth{Rares Vernica}{\hp\prevuci}
\auth{Jian Wen}{\oracle\prevucr}
\auth{Till Westmann}{\oracle}
\\[1ex]
\affaddr{
\uci University of California, Irvine 
\cloudera Cloudera Inc.
\google Google
\ibm IBM
\marklogic MarkLogic Corp.
\pivotal Pivotal Inc.
\\
\ucr University of California, Riverside
\hp HP Labs
\oracle Oracle Labs
\\[1ex]
\email{mjcarey@ics.uci.edu}
}
}

\date{\today}

\begin{document}
\maketitle

\begin{abstract}
AsterixDB is a new, full-function BDMS (Big Data Management System) with a feature set that distinguishes it from other platforms in today's open source Big Data ecosystem. 
Its features make it well-suited to applications like web data warehousing, social data storage and analysis, and other use cases related to Big Data. 
AsterixDB has a flexible NoSQL style data model; 
a query language that supports a wide range of queries;
a scalable runtime; 
partitioned, LSM-based data storage and indexing (including B\textsuperscript{+}-tree, R-tree, and text indexes); 
support for external as well as natively stored data; 
a rich set of built-in types; 
support for fuzzy, spatial, and temporal types and queries; 
a built-in notion of data feeds for ingestion of data; 
and transaction support akin to that of a NoSQL store.

Development of AsterixDB began in 2009 and led to a mid-2013 initial open source release. 
This paper is the first complete description of the resulting open source AsterixDB system. 
Covered herein are the system's data model, its query language, and its software architecture. 
Also included are a summary of the current status of the project and a first glimpse into how AsterixDB performs when compared to alternative technologies, including a parallel relational DBMS, a popular NoSQL store, and a popular Hadoop-based SQL data analytics platform, for things that both technologies can do. Also included is a brief description of some initial trials that the system has undergone and the lessons learned (and plans laid) based on those early "customer" engagements.
\end{abstract}

\footnotetext[1]{work done at the University of California, Irvine}
\footnotetext[4]{work done at the University of California, Riverside}

\section{Overview}

The Big Data era is upon us. 
A wealth of digital information is being generated daily through social networks, blogs, online communities, news sources, and mobile applications as well as from a variety of sources in our increasingly sensed surroundings. 
Organizations and researchers in most domains today recognize that tremendous value and insight can be gained by capturing this data and making it available for querying and analysis, and doing so is one of the major focuses of today's Big Data movement.  
Domains where the timely availability of information derived from Big Data could be tremendously beneficial include public safety, public health, national security, law enforcement, medicine, marketing, political science, and governmental policy-making, to name but a few.
But at least one "minor detail" remains: Where are we going to put all of this information, and just how is it going to be managed and accessed over time?

In 2009, the NSF-sponsored Asterix project set out to develop a next-generation system to ingest, manage, index, query, and analyze mass quantities of semi-structured data \cite{vision}. 
The project drew ideas from three areas -- semi-structured data management, parallel databases, and  first-generation Big Data platforms -- to create a next-generation, open-source software platform that scales by running on large, shared-nothing commodity computing clusters. 
The effort targeted a wide range of semi-structured use cases, ranging from ``data'' use cases -- whose data is well-typed and highly regular -- to ``content'' use cases -- whose data is irregular, involves more text, and whose schema may be hard to anticipate a priori or may never exist.
The initial results were released as an AsterixDB system \emph{beta} release in June of 2013. 
This paper aims to introduce AsterixDB to both the Big Data and database management system communities by providing a technical overview of its user model (data model and query language) and its software architecture.

To distinguish it from current Big Data analytics platforms, which query but don't store or manage data, we classify AsterixDB as a \emph{Big Data Management System} (BDMS).
One of the project's informal mottos is ``one size fits a bunch'', and we hope AsterixDB will prove useful for a wider range of use cases than are addressed by any one of the current Big Data technologies (e.g., Hadoop-based query platforms or key-value stores).
We aim to reduce the need for ``bubble gum and baling wire'' constructions involving multiple narrower systems and corresponding data transfers and transformations.
Our design decisions were influenced by what we believe are the key BDMS desiderata, namely:
\vspace{-1mm}
\begin{enumerate}\itemsep0pt
\item a flexible, semistructured data model for use cases ranging from ``schema first'' to ``schema never'';
\item a full query language with at least the power of SQL;
\item an efficient parallel query runtime;
\item support for data management and automatic indexing;
\item support for a wide range of query sizes, with processing cost proportional to the task at hand;
\item support for continuous data ingestion;
\item the ability to scale gracefully in order to manage and query large volumes of data by using large clusters;
\item support for today's common ``Big Data data types'', e.g., textural, temporal, and spatial data values.
\end{enumerate}
\vspace{-1mm}
AsterixDB aims to cover all of these, distinguishing it from existing data management technologies like Big Data analytics platforms (e.g. Hadoop-based query platforms---missing 4 and 5), NoSQL stores (missing 2), and parallel RDBMS (do not work well for 1).

\begin{figure}
\centering
\includegraphics[width=2.9in]{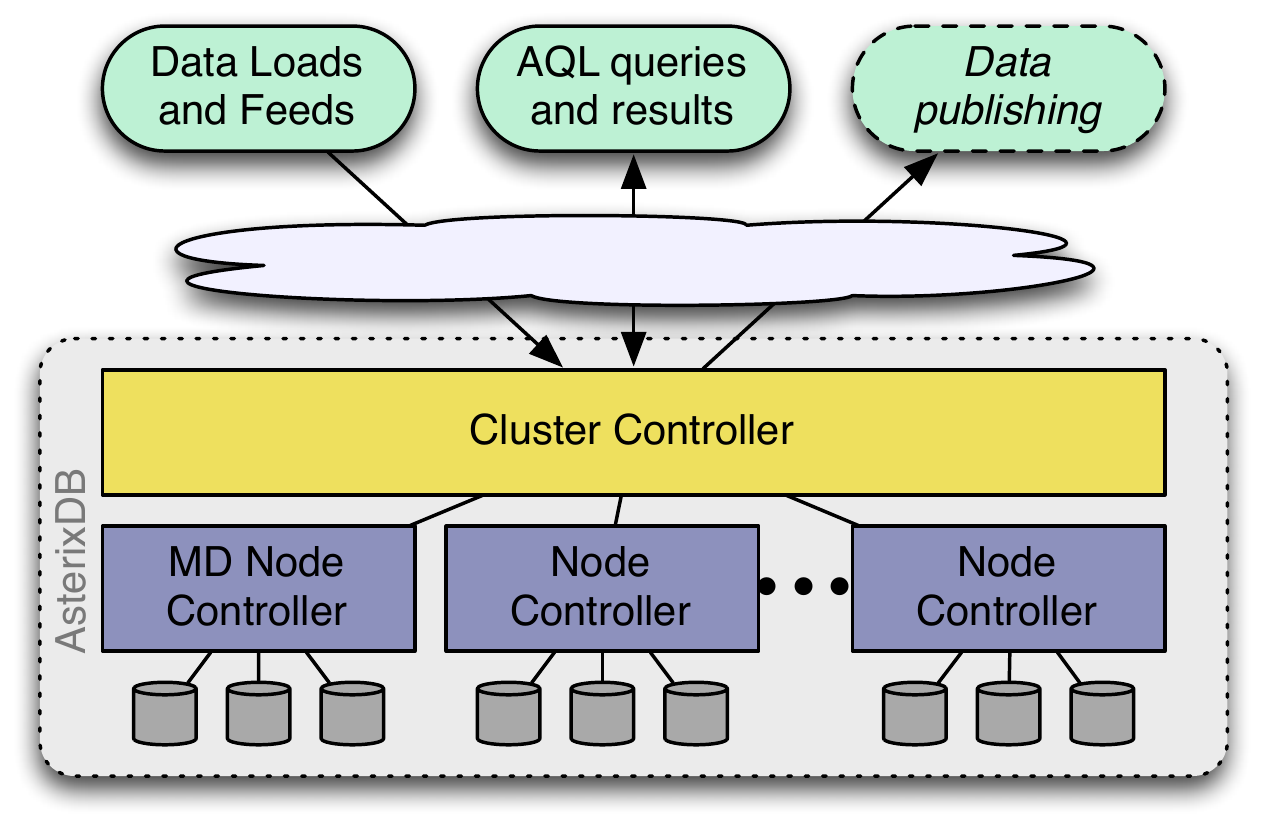}
\vspace*{\betweenfigureandcaption}
\caption{AsterixDB system overview\label{fig:marke}}
\vspace{-0.1in}
\end{figure}

Figure \ref{fig:marke} provides a high-level overview of AsterixDB and its logical architecture.  
Data enters through loading, continuous feeds, and/or insertion queries.
Data is accessed via queries and the return (synchronously or asynchronously) of their results. 
The Cluster Controller in Figure \ref{fig:marke} is the logical entry point for user requests; the Node Controllers and Metadata (MD) Node Controller provide access to AsterixDB's metadata and the aggregate processing power of the underlying shared-nothing cluster.
The figure's dotted \emph{Data publishing} path indicates that we are working towards eventually adding support for continuous queries and notifications.

The remainder of this paper is organized as follows: Section 2 covers the data definition side of AsterixDB, including its flexible JSON-based data model (ADM).
Section 3 covers data manipulation via AsterixDB's declarative query language (AQL). 
Section 4 describes the architecture of the system at the next level of detail, including its (openly) layered structure, its ingestion-targeted approach to data storage and indexing, its transactional support, and its data feed facility.
Section 5 discusses the current status of the AsterixDB code base and the open source R\&D effort, mentions some early use cases and their impact on the system, and shares a set of initial performance numbers to show how AsterixDB compares today to other available technologies in that regard. 
Section 6 provides a wrap-up and highlights the project's next steps.
\section{Data Definition}

In this section we describe AsterixDB's data definition features.  We illustrate them by example through a small and slightly silly scenario based on information about users and their messages from a popular (hypothetical) social network called Mugshot.com.

\subsection{Dataverses, Datatypes, and Datasets\label{ssec:dataverse}}

The top-level organizing concept in AsterixDB is the \emph{Dataverse}.
A Dataverse, short for "data universe", is a place (akin to a database in an RDBMS) within which one can create and manage the types, Datasets, functions, and other artifacts for a given application.
Initially, an AsterixDB instance contains no data other than the system catalogs, which live in a system-defined Dataverse (the Metadata Dataverse).
To store data in AsterixDB, one creates a Dataverse and then populates it with the desired Datatypes and Datasets.
A \emph{Datatype} specifies what its definer wants AsterixDB to know, a priori, about a kind of data that it will be asked to store.
A \emph{Dataset} is a stored collection of data instances of a Datatype, and one can define multiple Datasets of a given Datatype.
AsterixDB ensures that data inserted into a Dataset conforms to its specified type. 

Since AsterixDB targets semi-structured data, its data model provides the concept of \emph{open} (\emph{vs.~closed}) Datatypes.
When defining a type, the definer can use open Datatypes and tell AsterixDB as little or as much about their data up front as they wish.  
(The more AsterixDB knows about the potential residents of a Dataset, the less it needs to store in each individual data instance.)  
Instances of open Datatypes are allowed to have additional content, beyond what the type specifies, as long as they at least contain the information prescribed by the Datatype definition.  
Open types allow data to vary from one instance to another, leaving "wiggle room" for instance-level variations as well as for application evolution (in terms of what can be stored in the future). 
To restrict the objects in a Dataset to contain only what a Datatype says, with nothing extra in instances, one can opt to define a closed Datatype for that Dataset.
AsterixDB prevents users from storing objects with extra or illegally missing data in such a data set.
The closed, flat subset of ADM is thus relational.
Datatypes are open by default and closed only if their definition says so, the reason for this choice being that many Big Data analysts seem not to favor \emph{a priori} schema design and ADM's design targets semi-structured data.
To the best of our knowledge, this support for open and closed Datatypes is novel---we have not seen similarly flexible schema facilities in other data management systems, and it consistently garners very positive reactions when AsterixDB is presented to outside groups.

Shape-wise, ADM is a superset of JSON \cite{json}---ADM is what one gets by extending JSON with a larger set of Datatypes (e.g. datetime) and additional data modeling constructs (e.g. bags) drawn from object databases and then giving it a schema language.
We chose JSON for its self-describing nature, relative simplicity, and growing adoption in the Web world.
Note that, unlike AsterixDB, current JSON-based data platforms do not offer the option to define all (or part) of their data's schema.

\begin{ddl}[Dataverse and data types]{ddl:types}
drop dataverse TinySocial if exists;
create dataverse TinySocial;
use dataverse TinySocial;

create type EmploymentType as open {
    organization-name: string,
    start-date: date,
    end-date: date?
}

create type MugshotUserType as {
    id: int32,
    alias: string,
    name: string,
    user-since: datetime,
    address: {
        street: string,
        city: string,
        state: string,
        zip: string,
        country: string
    },
    friend-ids: {{ int32 }},
    employment: [EmploymentType]
}

create type MugshotMessageType as closed {
    message-id: int32,
    author-id: int32,
    timestamp: datetime,
    in-response-to: int32?,
    sender-location: point?,
    tags: {{ string }},
    message: string
}
\end{ddl}

We illustrate ADM by defining a Dataverse called \emph{TinySocial} to hold the Datatypes and Datasets for Mugshot.com.
The data definition statements in Data definition \ref{ddl:types} show how we can create the Dataverse and a set of ADM types to model Mugshot.com's users, their users' employment histories, and their messages. 
The first three lines tell AsterixDB to drop the old TinySocial Dataverse, if one exists, and to create a new Dataverse and make it the focus of the statements that follow. 
The first type creation statement creates a Datatype to hold information about one piece of a Mugshot user's employment history. 
It defines a record type with a mix of string and date data, much like a (flat) relational tuple.
Its first two fields are mandatory, with the last one (end-date) being optional as indicated by the "?" that follows it.
An optional field in ADM is like a nullable field in SQL -- it may be present or missing, but when present, its data type will conform to the Datatype's specification for it. Also, because EmploymentType is open, it is important to note that additional fields will be allowed at the instance level. 

\begin{table} [htbp]
\centering
\small
\begin{tabular}{|l|l|l}
\hline 
\textbf{Types}      & \textbf{functions}                     \\
\hline 
string              & contains                               \\
                    & like                                   \\
                    & matches                                \\
                    & replace                                \\
                    & word-tokens                            \\
                    & edit-distance                          \\
                    & edit-distance-check                    \\
                    & edit-distance-contains                 \\
\hline 
\emph{bag}          & similarity-jaccard                     \\
                    & similarity-jaccard-check               \\
\hline 
date/time/datetime  & current-date/time/datetime             \\
interval            & interval-start-from-date/time/datetime \\
duration            & adjust-datetime-for-timezone           \\
day-time-duration   & adjust-time-for-timezone               \\
year-month-duration & subtract-date/time/datetime            \\
                    & interval-bin                           \\
                    & \emph{Allen's relations} on intervals  \\\hline 
point               & spatial-distance                       \\
line                & spatial-area                           \\
rectangle           & spatial-intersect                      \\
circle              & spatial-cell                           \\
polygon             &                                        \\
\hline 
\end{tabular}
\caption{Sample of advanced types and functions\label{tab:types}}
\end{table}

The second create type statement creates a Datatype for Mugshot users.
MugshotUserType is also open since that is the default for AsterixDB Datatypes.
This second type highlights several additional features of ADM.
The address field illustrates one way that ADM is richer than the relational model; it holds a nested record containing the primary address of a user.
The friend-ids field shows another extension of ADM over the relational model and also over JSON.
This field holds a bag (unordered list) of integers -- presumably the Mugshot user ids for this user's friends.
Lastly for this type, its employment field is an ordered list of employment records, again a beyond-flat-relational structure.

The last create type statement in the example defines a Datatype to store the content of a Mugshot message.
In this case, since the type definition for Mugshot messages says \emph{closed}, the fields that it lists will be the only fields that instances of this type will be allowed to contain in Datasets of this type.
Also, among those fields, in-response-to and sender-location are optional, while the rest must be present in valid instances of the type.

Recall that AsterixDB aims to store and query not just Big Data, but Big \emph{Semi-structured} Data. 
Most of the fields in the create type statements above could be omitted, if desired, while changing only two things in terms of the example.  
One change would be the size of the data on disk. 
AsterixDB stores information about the fields defined \emph{a priori} as separate metadata; information about fields that are "just there" in instances of open Datatypes is stored within each instance.
A logical change would be that AsterixDB would be more flexible about the contents of records in the resulting Datasets, as it enforces only the specified details of the Datatype associated with a given Dataset.
The only fields that must currently be specified \emph{a priori} are the primary key fields. 
This restriction is temporary, as AsterixDB's next release will offer auto-generated keys.

One other important feature of AsterixDB for managing today's Big Data is its built-in support for useful advanced primitive types and functions, specifically those related to space, time, and text.
Table \ref{tab:types} lists some of the advanced ADM primitive types as well as some of their corresponding AQL functions.
A complete list and more details can be found in the AsterixDB documentation \cite{docs}.



\subsection{Dataset and Index Creation}\label{ss:dataset}

Having defined our Datatypes, we can now proceed to create a pair of Datasets to store the actual data. 

\begin{ddl}[Datasets and indexes]{ddl:datasets}
create dataset MugshotUsers(MugshotUserType)
    primary key id;
create dataset MugshotMessages(MugshotMessageType)
    primary key message-id;

create index msUserSinceIdx 
    on MugshotUsers(user-since);
create index msTimestampIdx
    on MugshotMessages(timestamp);
create index msAuthorIdx
    on MugshotMessages(author-id) type btree;
create index msSenderLocIndex
    on MugshotMessages(sender-location) type rtree;
create index msMessageIdx
    on MugshotMessages(message) type keyword;
\end{ddl}

\eat{
use dataverse TinySocial;

load dataset MugshotUsers using localfs
(("path"="127.0.0.1:///Users/tillw/Documents/Papers/AsterixDB/703942dkdpdw/data/msu.adm.txt"),("format"="adm"));

load dataset MugshotMessages using localfs
(("path"="127.0.0.1:///Users/tillw/Documents/Papers/AsterixDB/703942dkdpdw/data/msm.adm.txt"),("format"="adm"));
}
\eat{
use dataverse TinySocial;

for $u in dataset MugshotUsers return $u;
for $m in dataset MugshotMessages return $m;
}

The two ADM DDL statements shown in Data definition \ref{ddl:datasets} will create Datasets to hold data in the TinySocial Dataverse.
The first creates a Dataset called MugshotUsers; 
this Dataset will store data conforming to MugshotUserType and has the id field as its primary key. 
The primary key is used by AsterixDB to uniquely identify instances for later lookup and for use in secondary indexes.
Each AsterixDB Dataset is stored (and indexed) as a B\textsuperscript{+}-tree keyed on primary key; secondary indexes point to indexed data by its primary key. 
Also, in an AsterixDB cluster, the primary key is used to hash-partition (shard) the Dataset across the cluster's nodes. 
The ^create dataset^ statement for MugshotMessages is similar. 


The two ^create dataset^ statements are followed by five more DDL statements, each requesting the creation of a secondary index on a field of one of the Datasets. 
The first two will index MugshotUsers and MugshotMessages on their user-since and timestamp fields. 
These indexes will be B\textsuperscript{+}-tree indexes, as their type is unspecified and ^btree^ is the default. 
The other three show how to explicitly specify the desired index type. 
In addition to ^btree^, ^rtree^ and inverted ^keyword^ indexes are supported. 
Indexes can have composite keys, and more advanced text indexing is also available (^ngram(k)^, where k is the gram length, for fuzzy searching).

\subsection{External Data}

So far we have explained how to specify Datatypes, Datasets, and indexes for AsterixDB to store and manage in its role as a full BDMS.  
AsterixDB also supports direct access to externally resident data.
Data does not need to be pre-loaded and handed to AsterixDB for storage and management just to be queried---avoiding the costly (and often infeasible) transfer or duplication of ``Big Data''.
The current AsterixDB system offers external data adaptors to access local files that reside on the Node Controller nodes of an AsterixDB cluster and to access data residing in HDFS. 
To illustrate, Data definition \ref{ddl:ext} shows the definition of an external Dataset based on a local file.  
In this case, the local file is a CSV version (Figure \ref{fig:csv-log}) of an Apache log file (Figure \ref{fig:apache-log}). 
When accessed at query time, CSV parsing of the data into ADM object instances will be driven by the type definition associated with the Dataset.

Once defined in this manner, an external Dataset can be accessed in a query just like any internal Dataset.
Unlike an internal Dataset, however, external Datasets in the current release of AsterixDB are limited to being read-only and static and indexes cannot be created on them.
(Support for incrementally refreshable external data sets, as well as indexing externally resident data, is being added to the system and will arrive in a mid-summer 2014 AsterixDB release.)


\begin{figure*}
\begin{lstlisting}
12.34.56.78 - Nicholas [22/Dec/2013:12:13:32 -0800] "GET / HTTP/1.1" 200 2279 
12.34.56.78 - Nicholas [22/Dec/2013:12:13:33 -0800] "GET /list HTTP/1.1" 200 5299 
\end{lstlisting}
\vspace*{\betweenfigureandcaption}
\caption{Apache HTTP server common log format\label{fig:apache-log}}
\end{figure*}

\begin{figure*}
\begin{lstlisting}
12.34.56.78|2013-12-22T12:13:32-0800|Nicholas|GET|/|200|2279 
12.34.56.78|2013-12-22T12:13:33-0800|Nicholas|GET|/list|200|5299 
\end{lstlisting}
\vspace*{\betweenfigureandcaption}
\caption{CSV version of web server log\label{fig:csv-log}}
\end{figure*}

\begin{ddl}[External data]{ddl:ext}
create type AccessLogType as closed {
    ip: string,
    time: string,
    user: string,
    verb: string,
    path: string,
    stat: int32,
    size: int32
}

create external dataset AccessLog(AccessLogType)
    using localfs
        (("path"="{hostname}://{path}"),
         ("format"="delimited-text"),
         ("delimiter"="|"));
\end{ddl}


\vspace{-0.1in}
\subsection{Data Feed Creation}\label{ss:feeds}

\emph{Data Feeds} are a built-in mechanism that AsterixDB offers to allow new data to be continuously ingested into one or more Datasets from external sources, incrementally populating the Datasets and their associated indexes.
Feed support is provided because the need to persist and index "fast flowing" data is ubiquitous in the Big Data world, and it otherwise involves gluing together multiple systems. 
We feel that, just as current DBMSs were created to provide the commonly required functionality to support data-centric applications, a BDMS should provide support for continuous data ingestion and should be responsible for managing the performance and fault-tolerance of the ingestion pipeline.

An AsterixDB data feed ingests a continuous stream of data into a Dataset. 
User queries then work against the stored data, not on the incoming stream, just as if the Dataset's contents had arrived via loading or insertions. 
With this approach, the system does not require a separate notion of queryable streams (with their different semantics, etc.) distinct from its support for stored data sets.

Data definition \ref{ddl:feed} shows the declaration of a Data Feed and the connecting of it to a Dataset for storage. 
A socket-based feed adaptor is used, allowing data from outside to be pushed at AsterixDB via a TCP/IP socket where the adaptor will listen for data.

\begin{ddl}[Data feeds]{ddl:feed}
use dataverse TinySocial;

create feed socket_feed
    using socket_adaptor 
        (("sockets"="{address}:{port}"),
         ("addressType"="IP"),
         ("type-name"="MugshotMessageType"),
         ("format"="adm"));

connect feed socket_feed to dataset MugshotMessages;
\end{ddl}
\eat{
("sockets"="127.0.0.1:9009")
}

In addition to the ^socket_adaptor^ there are a few built-in adaptors included with AsterixDB.
To customize an existing adaptor it is also possible to apply a previously defined function (see Section \ref{ss:udfs}) to the output of the adaptor.
Finally, AsterixDB also provides a mechanism to add custom adaptors to the system.

Feeds that process external data, like the one above, are called \emph{Primary Feeds}. 
AsterixDB also supports \emph{Secondary Feeds} that are fed from other feeds.
Secondary Feeds can be used, just like Primary Feeds, to transform data and to feed Datasets or feed other feeds.
More about the user model for feeds, its extensibility, and its implementation and performance can be found in \cite{AsterixFeeds}.  (Note: Data feeds are a "hidden" feature in the current open source system, as they have not yet been officially documented/released for general external use. That will change with the next release of AsterixDB, where Data Feed support will be the ``flagship'' new feature.)

\subsection{User Defined Functions}\label{ss:udfs}

One final DDL feature that should be mentioned is  support for reusable \emph{user-defined functions} (UDFs), which are similar in nature to views in relational databases.
(AsterixDB's AQL UDFs are essentially views with parameters.)
As the definition of such a function requires an AQL function body, we will provide an example in Query \ref{q:function} in Section \ref{sec:dml} and will provide more information about UDFs once we have introduced the reader to the basics of AQL.
\section{Data Manipulation}\label{sec:dml}

The query language for AsterixDB is AQL (Asterix Query Language).
Given the nature of ADM, we needed a language capable of dealing nicely with nesting and a lack of a priori schemas; we also wanted its semantics to be the same with and without schemas.
XQuery~\cite{xquery} had similar requirements from XML, so we chose to base AQL loosely on XQuery. 
ADM is simpler than XML, and XPath compatibility was irrelevant, so we jettisoned the ``XML cruft'' from XQuery---document order and identity, elements \emph{vs.} attributes, blurring of atomic and sequence values---keeping its core structure and approach to handling missing information.
We could have started over, but XQuery was co-designed by a diverse band of experienced language designers (SQL, functional programming, and XML experts) and we wanted to avoid revisiting many of the same issues and making mistakes that the W3C XQuery team had made, identified, and fixed in their years of work.
Starting from SQL would have been messier, syntactically and semantically, as ANSI SQL was designed for flat data---subqueries often have scalar-at-runtime-else-error semantics---and its treatment of nesting for its nested table feature is ugly and complex.
We wanted a much cleaner start for AQL.
(Also, since AQL is not SQL-based, AsterixDB is ``NoSQL compliant''.)

\eat{ 
AQL is an expression language; as such, the expression 1+1 is a valid AQL query that evaluates to 2.  
Most useful AQL queries are based on the FLWOR (``flower'') expression structure that AQL borrows from XQuery. 
The FLWOR syntax supports both the incremental binding (^for^) of variables to ADM data instances in a Dataset--or in the result of any AQL expression, actually--and the full binding (^let^) of variables to entire intermediate result expressions. 
FLWOR stands for ^for-let-where-order by-return^, naming the five most frequently used clauses of AQL's query syntax. 
AQL also has ^group by^ and ^limit^ clauses, as we will soon see. 
Roughly speaking, the ^for^ clause in AQL is like the ^FROM^ clause in SQL, the ^return^ clause in AQL is like the ^SELECT^ clause in SQL (but goes at the end instead of the start of a query), the ^let^ clause in AQL is like SQL’s ^WITH^ clause, and the ^where^ and ^order by^ clauses in both languages are similar.
}

AQL is an expression language; as such, the expression 1+1 is a valid AQL query that evaluates to 2.  
Most useful AQL queries are based on the FLWOR (\flwor) expression structure that AQL borrows from XQuery. 
FLWOR stands for ^for^-^let^-^where^-^order by^-^return^, naming the five most frequently used clauses of the full AQL query syntax. 
A ^for^ clause provides an incremental binding of ADM instances in a sequence (e.g. a Dataset) to variables, while the ^let^ clause provides a full binding of variables to entire intermediate result expressions.
Roughly speaking, the ^for^ clause in AQL is like the ^FROM^ clause in SQL, the ^return^ clause in AQL is like the ^SELECT^ clause in SQL (but goes at the end of a query), the ^let^ clause in AQL is like SQL's ^WITH^ clause, and the ^where^ and ^order by^ clauses in both languages are similar.
AQL also has ^group by^ and ^limit^ clauses, as we will see shortly. 

We will describe AQL by presenting a series of illustrative AQL queries based on our Mugshot.com schema. 
Most of the salient features of AQL will be presented, and of course more detail can be found in the online AsterixDB documentation \cite{docs}.

\begin{query}[All Datasets and indexes in the system]{q:metadata}
for $ds in dataset Metadata.Dataset return $ds;
for $ix in dataset Metadata.Index return $ix;
\end{query}

We begin our AQL tour with Query \ref{q:metadata}, which shows two queries that use the simplest (useful) ^for^ and ^return^ clauses and show how the keyword ^dataset^ is used to access an AsterixDB Dataset in AQL.  
The queries each return the instances in a target Dataset, and each targets a Dataset in the Metadata Dataverse. 
The first one returns the set of all Datasets that have been defined so far, and the second one returns the set of all known indexes. 
These queries highlight the useful fact that AsterixDB ``eats its own dog food'' with respect to system catalogs---AsterixDB metadata is AsterixDB data, so (unlike Hive, for example) AsterixDB catalogs are stored in the system itself, as is also true in most RDBMSs.

\eat{ 
\begin{query}[Find a record]{q:lookup}
for $user in dataset MugshotUsers
where $user.id = 8
return $user;
\end{query}

Query \ref{q:lookup} illustrates an AQL where clause. This query's ^for^ clause conceptually iterates over all records in the Dataset ^MugshotUsers^, binding each to the variable ^$user^, returning ADM records whose ^id^ field contains the value 8.
(Of course, as ^id^ is the key field of the Dataset ^MugshotUsers^, the query's actual evaluation will be more efficient, utilizing the Dataset's primary index under the covers.) 
}

\begin{query}[Datetime range scan]{q:range-scan}
for $user in dataset MugshotUsers
where $user.user-since >= datetime('2010-07-22T00:00:00')
  and $user.user-since <= datetime('2012-07-29T23:59:59')
return $user;
\end{query}

Query \ref{q:range-scan} illustrates an AQL where clause. This query's ^for^ clause conceptually iterates over all records in the Dataset ^MugshotUsers^, binding each to the variable ^$user^, returning the ADM records for users  that joined mugshot.com from July 22, 2010 to July 29, 2012.
(Of course,  the query's actual evaluation may be more efficient, e.g., using an index on ^user-since^ under the covers.) 

\begin{query}[Equijoin]{q:equijoin}
for $user in dataset MugshotUsers
for $message in dataset MugshotMessages
where $message.author-id = $user.id
  and $user.user-since >= datetime('2010-07-22T00:00:00')
  and $user.user-since <= datetime('2012-07-29T23:59:59')
return {
  "uname" : $user.name,
  "message" : $message.message
};
\end{query}

Query \ref{q:equijoin} shows a first example where new records are being synthesized by a query.
It first selects user records like Query \ref{q:range-scan}, but then it also selects matching message records whose ^author-id^ is equal to the user's ^id^---an equijoin expressed in AQL.
For each match, it creates a new ADM record containing two fields, ^uname^ and ^message^, that will contain the user’s name and the message text, respectively, for that user/message pair. 
Note that the query returns a sequence of flat records, i.e., it repeats the user name with every message. Also, as the match predicate is only true when a match exists, users without matching messages and messages without matching users are not returned.

\begin{query}[Nested left outer-join]{q:outer-join}
for $user in dataset MugshotUsers
where $user.user-since >= datetime('2010-07-22T00:00:00')
  and $user.user-since <= datetime('2012-07-29T23:59:59')
return {
  "uname" : $user.name,
  "messages" :
    for $message in dataset MugshotMessages
    where $message.author-id = $user.id
    return $message.message
};
\end{query}

Query \ref{q:outer-join} shows a more natural way to match users and messages in AQL.
Users of interest are targeted by the first ^for^ and ^where^ clause, and a nested ^FLWOR^ expression synthesizes a bag of matching messages for each user.
In contrast to Query \ref{q:equijoin}, the result will include users who have not yet sent any messages, and the result will be a set of nested ADM records, one per user, with each user's messages listed ``inside'' their user record.
This is the equivalent of a SQL left outer join in AQL, but with a more natural result shape since ADM permits nesting (unlike the relational model).

\begin{query}[Spatial join]{q:spatial-join}
for $t in dataset MugshotMessages
return {
  "message" : $t.message,
  "nearby-messages":
    for $t2 in dataset MugshotMessages
    where spatial-distance($t.sender-location,
                           $t2.sender-location) <= 1
    return { "msgtxt" : $t2.message }
};
\end{query}

Query \ref{q:spatial-join} shows another ``join'' example, one that illustrates the use of AsterixDB's spatial data support. 
This query goes through the set of all Mugshot messages and, for each one, uses a nested query to pair it with a bag of messages sent from nearby locations.

\begin{query}[Fuzzy selection]{q:fuzzy-string-join}
set simfunction "edit-distance";
set simthreshold "3";

for $msu in dataset MugshotUsers
for $msm in dataset MugshotMessages
where $msu.id = $msm.author-id
  and (some $word in word-tokens($msm.message)
       satisfies $word ~= "tonight")
return {
  "name" : $msu.name,
  "message" : $msm.message
};
\end{query}

Query \ref{q:fuzzy-string-join} illustrates some of AsterixDB's \emph{fuzzy} matching capabilities.
This query returns the sending user name along with a message if one of the words in a tokenization of the message fuzzily matches (^~=^) "tonight".
\emph{Fuzzy} in this case means that the edit-distance is less than or equal to 3, e.g., a message that contains the word "tonite" would also be returned. 
The ^set^ statements in the query's prologue are used to specify the desired fuzzy matching semantics;  a functional syntax for fuzzy matching is also available to specify the matching semantics within the matching predicate itself (which is needed if a query requires multiple fuzzy match predicates, each with different match semantics).

\begin{query}[Existential quantification]{q:existential}
for $msu in dataset MugshotUsers
where (some $e in $msu.employment 
       satisfies is-null($e.end-date)
         and $e.job-kind = "part-time")
return $msu;
\end{query}

Query \ref{q:existential} illustrates two more advanced aspects of AQL, namely existential quantification and the use of an open field (one that's not part of the type definition for the data in question).
This example query uses existential quantification in its ^where^ clause to find users who have a current employment record (i.e., one with a null ^end-date^) that has a job-kind field whose value is ^"part-time"^.
(Note that ^job-kind^ is not declared to be a field of ^EmploymentType^.)

\begin{query}[Universal quantification and function definition]{q:function}
create function unemployed() {
  for $msu in dataset MugshotUsers
  where (every $e in $msu.employment 
         satisfies not(is-null($e.end-date)))
  return {
    "name" : $msu.name,
    "address" : $msu.address
  }
};
\end{query}

\begin{query}[Function use]{q:call}
for $un in unemployed()
where $un.address.zip = "98765"
return $un
\end{query}

Query \ref{q:function} defines a function (similar to a view in SQL) that returns the name and address of unemployed users.
It tests for unemployed users by seeing that all their employments have ended.
Query \ref{q:call} then uses this function and selects all unemployed users in the ZIP code 98765. Such a function can be written by an experienced user (one with a taste for universal quantifiers) and then used by a novice user (one with more normal tastes).

\begin{query}[Simple aggregation]{q:agg}
avg(
  for $m in dataset MugshotMessages 
  where $m.timestamp >= datetime("2014-01-01T00:00:00")
    and $m.timestamp <  datetime("2014-04-01T00:00:00")
  return string-length($m.message)
)
\end{query}

Like any reasonably expressive query language, AQL includes support for aggregation. Query \ref{q:agg} is a first example of an AQL aggregate query, computing the average message length during a time interval of interest. 
AQL aggregates include ^count^, ^min^, ^max^, ^avg^, and ^sum^ as well as ^sql-count^, ^sql-min^, ^sql-max^, ^sql-avg^, and ^sql-sum^. AQL's own aggregates have what we consider to be ``proper'' semantics regarding null values; e.g., the average of a set of values is null (unknown) if any of the values encountered is null. 
AQL also offers a set of aggregate functions with SQL's ``best guess'' null handling semantics, wherein the average of a set of values with nulls is the sum of the non-null values divided by the number of non-null values.

\eat{
\begin{query}[Grouping and aggregation]{q:gagg}
for $msg in dataset MugshotMessages
group by $aid := $msg.author-id with $msg
return {
  "author" : $aid,
  "no messages" : count($msg)
};
\end{query}
}

\begin{query}[Grouping with sorting and limits]{q:sort}
for $msg in dataset MugshotMessages
where $msg.timestamp >= datetime("2014-02-20T00:00:00")
  and $msg.timestamp <  datetime("2014-02-21T00:00:00")
group by $aid := $msg.author-id with $msg
let $cnt := count($msg)
order by $cnt desc
limit 3
return {
  "author" : $aid,
  "no messages" : $cnt
};
\end{query}

AQL supports grouped aggregation as well and -- since Big Data often means ``many groups'' -- also provides the machinery to get the `top'' results, not all results.
Query \ref{q:sort} illustrates how the ^group^ ^by^, ^order^ ^by^, and ^limit^ clauses can be used to group and count messages by their sender and to report the results only for the three chattiest Mugshot.com users.

\begin{query}[Active users]{q:active}
let $end := current-datetime()
let $start := $end - duration("P30D")
for $user in dataset MugshotUsers
where some $logrecord in dataset AccessLog
  satisfies $user.alias = $logrecord.user
  and datetime($logrecord.time) >= $start 
  and datetime($logrecord.time) <= $end
group by $country := $user.address.country with $user
return {
  "country" : $country,
  "active users" : count($user)
}
\end{query}

Query \ref{q:active} identifies all \emph{active} users and then counts them grouped by country.
The query considers users that had activity in the last 30 days to be active.
Activity data is taken from the web server logs that are exposed as an external dataset (see Figure \ref{fig:csv-log}).
This example also shows the use of datetime arithmetic in AQL.

\begin{query}[Left outer fuzzy join]{q:fuzzy-join}
set simfunction "jaccard";
set simthreshold "0.3";

for $msg in dataset MugshotMessages
let $msgsSimilarTags := (
  for $m2 in dataset MugshotMessages
    where  $m2.tags ~= $msg.tags
      and $m2.message-id != $msg.message-id
    return $m2.message
  )
where count($msgsSimilarTags) > 0
return {
  "message" : $msg.message,
  "similarly tagged" : $msgsSimilarTags    
};
\end{query}

Last but not least, Query \ref{q:fuzzy-join} closes out our tour of AQL's query power by showing how one can express fuzzy joins in AQL. 
This example analyzes Mugshot.com's messages by finding, for each message where there are one or more counterparts with similar tags, the similarly tagged messages. 
Here similarity means Jaccard similarity of 0.3 (messages with more than 30\% of the same tags). 
Under the covers AsterixDB has built-in support for both ad hoc parallel fuzzy joins as well as indexed fuzzy joins.

\begin{update}[Simple insert]{upd:insert}
insert into dataset MugshotUsers
(
  { 
    "id":11, 
    "alias":"John", 
    "name":"JohnDoe", 
    "address":{ 
      "street":"789 Jane St", 
      "city":"San Harry", 
      "zip":"98767", 
      "state":"CA", 
      "country":"USA" 
    }, 
    "user-since":datetime("2010-08-15T08:10:00"), 
    "friend-ids":{{ 5, 9, 11 }}, 
    "employment":[{ 
        "organization-name":"Kongreen", 
        "start-date":date("2012-06-05")
    }]
  }
);
\end{update}

\begin{update}[Simple delete]{upd:delete}
delete $user from dataset MugshotUsers 
where $user.id = 11;
\end{update}

Data can enter AsterixDB via loading, feeds, or insertion. AQL's support for ^insert^ operations is illustrated in Update \ref{upd:insert}; its corresponding support for doing ^delete^ operations is shown in Update \ref{upd:delete}. 
The data to be inserted is specified as any valid AQL expression; in this case the expression is a new record whose content is known a priori. 
Delete operations' ^where^ clauses can involve any valid AQL boolean expression. 
Currently the AsterixDB answer for modifying data in a Dataset is ``out with the old, in with the new\!''---i.e., a delete followed by an insert. 
(We are targeting append-heavy use cases initially, not modification-heavy scenarios.)

In terms of the offered degree of transaction support, AsterixDB supports record-level ACID transactions that begin and terminate implicitly for each record inserted, deleted, or searched while a given AQL statement is being executed. 
This is quite similar to the level of transaction support found in today’s NoSQL stores. 
AsterixDB does not support multi-statement transactions, and in fact an AQL statement that involves multiple records can itself involve multiple independent record-level transactions \cite{docs}.
\section{System Architecture}\label{sec:sysarch}

\begin{figure}
\centering
\includegraphics[width=2.7in]{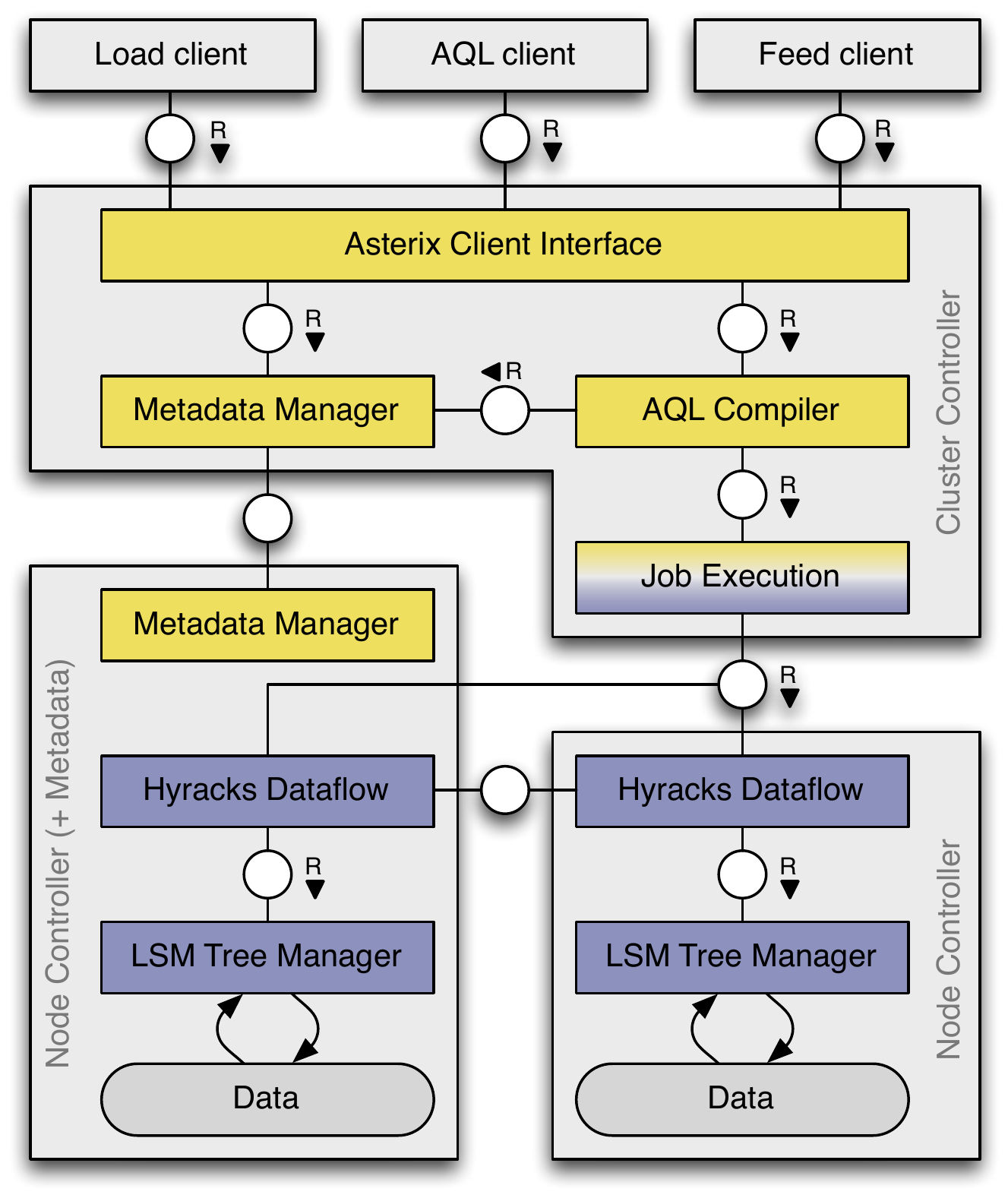}
\vspace*{\betweenfigureandcaption}
\caption{Architecture\label{fig:arch}}
\vspace{-0.1in}
\end{figure}

Figure \ref{fig:arch} contains an FMC diagram \cite{FMC} showing the next level of detail from Figure \ref{fig:marke}'s AsterixDB architectural summary.  
The top box shows the components of a Query Control Node, which in the current release coincides with the Hyracks Cluster Controller.  

This node receives queries via an HTTP-based API and returns their results to the requester either synchronously or asynchronously (in which case a handle to the result is returned and the client can inquire about the query's status and request the result via the handle).
The Query Control Node also runs (a) the AQL compiler that translates AQL statements to Job Descriptions for the dataflow-engine Hyracks and (b) the Job Executor that distributes the Job Descriptions to the Hyracks Node Controllers and starts the execution.
The Worker Nodes -- represented by the boxes at the bottom of Figure \ref{fig:arch} -- have to (a) manage their partitions of the data stored in LSM-trees and to (b) run their parts of each Hyracks Job.
In the following subsection we will first describe Hyracks and Algebricks (which is a core part of the AQL compiler), before discussing a few details on storage and indexing, transactions and data feeds in AsterixDB. 

\begin{figure}[htb]
\centering
\includegraphics[width=3.0in]{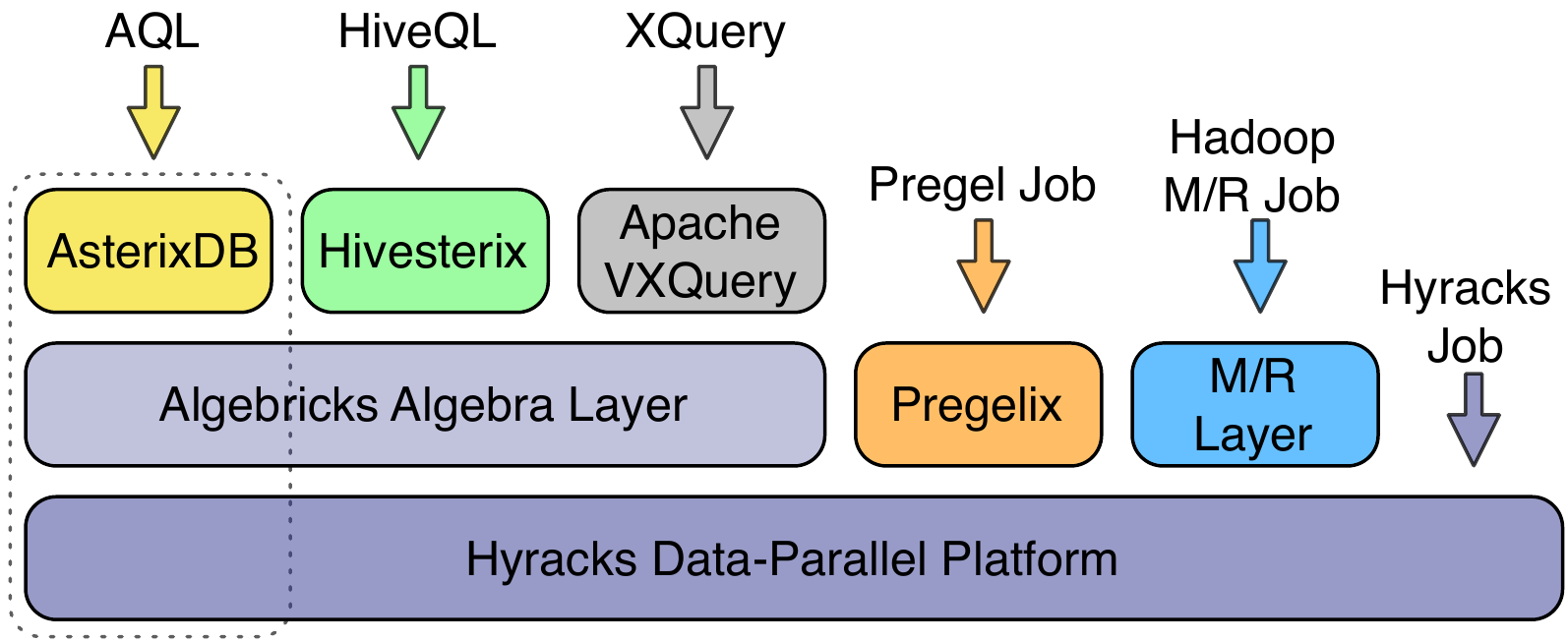}
\vspace*{\betweenfigureandcaption}
\caption{Asterix software stack\label{fig:stack}}
\vspace{-0.1in}
\end{figure}

\subsection{Hyracks}

The Hyracks layer of AsterixDB is the bottom-most layer of the Asterix software stack shown in Figure \ref{fig:stack}.
Hyracks is the runtime layer (\emph{a.k.a.}~executor) whose responsibility is to accept and manage data-parallel computations requested either by one of the layers above it in the Asterix software stack or potentially by direct end-users of Hyracks. 
In the case of AsterixDB, Hyracks serves as the scalable runtime engine for AQL queries once they have been compiled down into Hyracks Jobs for execution.

Jobs are submitted to Hyracks in the form of DAGs made up of \emph{Operators} and \emph{Connectors}. 
In Hyracks, Operators are responsible for consuming partitions of their inputs and producing output partitions. 
Connectors redistribute data from the output partitions and provide input partitions for the next Operator.
Hyracks currently provides a library of 53 Operators and 6 Connectors.
Operators include different Join Operators (HybridHash, GraceHash, NestedLoop), different Aggregation Operators (HashGroup and PreclusteredGroup), and a number of Operators to manage the lifecycle (create, insert, search, ...) of the supported index structures (B+-Tree, R-Tree, InvertedIndex, ...).
The Connectors are OneToOne, MToNReplicating, MToNPartitioning, LocalityAwareMToNPartitioning, MToNPartitioningMerging and HashPartitioningShuffle. 

\begin{figure}[htb]
\centering
\includegraphics[width=3.0in]{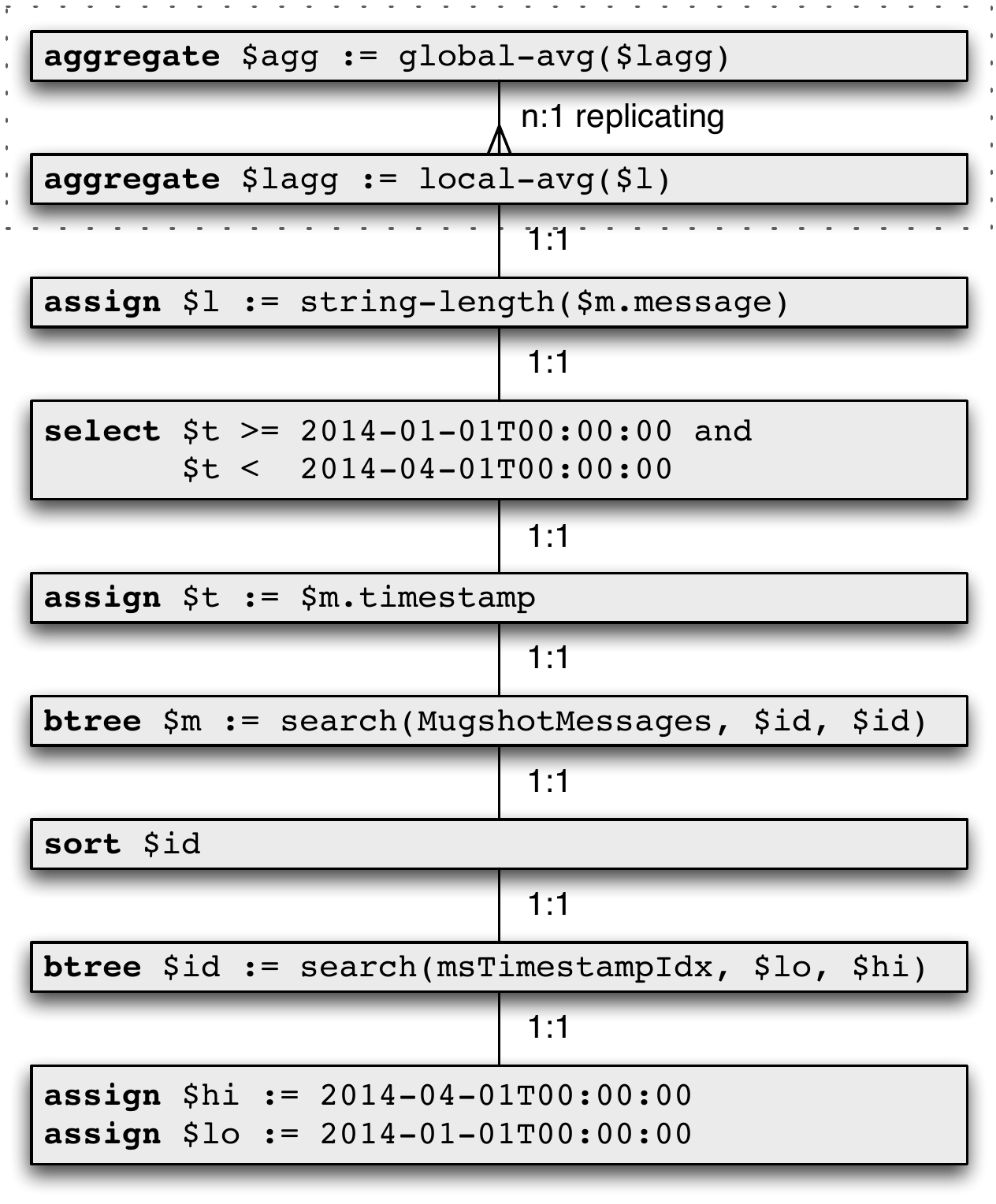}
\vspace*{\betweenfigureandcaption}
\caption{Hyracks job\label{fig:job}}
\vspace{-0.1in}
\end{figure}

Figure \ref{fig:job} depicts the Hyracks Job for Query \ref{q:agg}.
The boxes represent its Operators and the lines represent Connectors.
One thing that we see is that all Connectors, except for the last one, are OneToOneConnectors. 
This means that no redistribution of data is required for this part of the plan---and it can be evaluated on the node that stores the data.
The degree of parallelism (the number of Operator instances evaluating in parallel) for these Operators is the number of partitions that is used to store the Dataset.
Looking at the Operators bottom-up, we see that the first 2 Operators perform a search on a secondary index.
This will return a set of primary key values that are fed into the search Operator on the primary index.
Note that the primary keys are sorted first to improve the access pattern on the primary index.
Above the primary index access we (surprisingly) see two Operators that evaluate a predicate that should intuitively always be true for records that were identified by the search on the secondary index.
Section \ref{ss:tx} explains why this might not be the case and why this selection is needed.
Finally, we see that the ^avg^ function that encapsulates the rest of the query has been split into two Operators: a \emph{Local Aggregation Operator} that pre-aggregates the records for the local node and a \emph{Global Aggregation Operator} that aggregates the results of the the Local Aggregation Operators.
Between these Operators is a MToNReplicatingConnector.
As the degree of parallelism for the Global Aggregation Operator is constrained to be 1, the Connector replicates the results of all instances of the Local Aggregation Operators to the single instance of the Global Aggregation Operator.
This split maximizes the distributed computation and minimizes network traffic.

As the first step in the execution of a submitted Hyracks Job, its Operators are expanded into their constituent \emph{Activities}. 
While many Operators have a single Activity, some Operators consist of two or more Activities.
For example, a HybridHash Join Operator is made up of two Activities, the Join Build Activity and the Join Probe Activity. 
This expansion is made possible by APIs that Hyracks provides to Operator implementors to enable them to describe such behavioral aspects of an Operator.
Although Hyracks does not understand the specifics of the various activities of an Operator, exposing the blocking characteristics of Operators provides important scheduling information to Hyracks -- the separation of an Operator into two or more Activities surfaces the constraint that it can produce no output until all of its input has been consumed.

To execute a Job, Hyracks analyzes the Activity graph produced by the expansion described above to identify the collections of Activities (Stages) that can be executed at any time (while adhering to the blocking requirements of the Operators). 
Each Stage is parallelized and the resulting graph is executed in the order of the dependencies. 
More details about Hyracks' computational model, as well as its implementation and performance, are available in \cite{hyracks}. 

It is worth noting that Hyracks has been scale-tested on a 180-node cluster at Yahoo! Research
\cite{DBLP:journals/corr/abs-1203-0160}.
Each node had 2 quad-core Intel Xeon E5420 processors, leading to a total of 1440 cores across the cluster. 
Hyracks has been driven by Hivesterix \cite{hivesterix} to run a TPC-H workload on 90 nodes and by Pregelix \cite{pregelix} to run PageRank on 146 nodes.\footnote{Our collaboration with Yahoo! Research has ended and we do not have access to this cluster anymore. As a result we were not able to run the full AsterixDB stack at this scale.}
The initial experience with running these workloads led to a number of improvements that enabled Hyracks to run well at this scale. 
The most notable impact was a re-im\-ple\-men\-ta\-tion of the network layer to ensure that there is at most one TCP connection between any two given nodes at any point in time. 

\subsection{Algebricks}

Figure \ref{fig:stack} shows that the open-source Asterix software stack supports the AsterixDB system but also aims to address other Big Data requirements. 
To process a query, AsterixDB compiles an AQL query into an Algebricks algebraic program. 
This program is then optimized via algebraic rewrite rules that reorder the Algebricks Operators and introduce partitioned parallelism for scalable execution. 
After optimization code generation translates the resulting physical query plan into a corresponding Hyracks Job that uses Hyracks to compute the desired query result.  
The left-hand side of Figure \ref{fig:stack} shows this layering. 
As indicated there, the Algebricks algebra layer is actually data-model-neutral and it supports other high-level data languages on this software stack as well \cite{compiler13}.
Other languages today include Hivesterix \cite{hivesterix}, a Hive port that was built at UC Irvine, and Apache VXQuery, a parallel XQuery project that is incubating in the Apache open source world \cite{VXQuery}.

As indicated at the second layer of Figure \ref{fig:stack}, the Asterix open source software stack also offers a compatibility layer for users who have Hadoop MapReduce jobs \cite{hadoop:website} and wish to run them using the Hyracks runtime. 
It also includes other experimental Big Data programming models, most notably Pregelix \cite{pregelix}, an open source implementation of Google's Pregel programming model \cite{Pregel}.

\subsection{Storage and Indexing}

AsterixDB is novel in having wholly embraced Log-Structured Merge (LSM) trees \cite{lsm} as the underlying technology for all of its internal data storage and indexing. 
In LSM-trees the state of the index is stored in different components with different lifecycles.
Entries being inserted into an LSM-tree are placed into a component that resides in main memory---an \textit{in-memory component}. When
the memory occupancy of the in-memory component exceeds a specified threshold,
the entries are \textit{flushed} into a component of the index that resides on
disk---a \textit{disk component}. 
As components begin to accumulate on disk, they
are periodically merged with older disk components subject to some
\textit{merge policy} that decides when and what to merge.
As entries are initially stored in memory and are moved to persistent storage in bulk, LSM-trees avoid costly random disk I/O and, as a result, enable AsterixDB to support high ingestion rates---especially for continuous ingestion.

AsterixDB employs a software framework that enables ``LSM-ification'' of any kind of index structure that supports a specified primitive set of index operations. 
This framework converts an in-place update, disk-based data structure to a deferred-update, append-only data structure, enabling efficient ingestion into the index.
The current release of AsterixDB supports LSM-ified B\textsuperscript{+}-trees, R-trees, and inverted keyword and n-gram indexes.

As mentioned in Section \ref{ss:dataset}, Datasets are represented as partitioned LSM-based B\textsuperscript{+}-trees using hash-based partitioning on the Dataset's primary key.
Secondary indexes on a Dataset are node-local, i.e., their partitions refer to data in the primary index partition on the same storage node---enabling secondary index lookups without an additional network hop. 
Consequently, secondary index lookups must be routed to all of a Dataset's partition storage nodes, as the matching data could be in any partition. 
The lookups occur in parallel on all partitions.
The result of a secondary key lookup is a set of primary keys.
The resulting primary keys are used to look up the data itself. 
An example can be seen in Figure \ref{fig:job}, where the result of the B\textsuperscript{+}-tree search in ^msTimestampIdx^ is sorted before being used as an input to the search in ^MugshotMessages^. More details on native storage and indexing in AsterixDB are in \cite{AsterixLSM}.

As mentioned earlier, AsterixDB also provides external Dataset support so that such data, e.g., living in HDFS, does not need to be loaded into AsterixDB to be queryable. 
To do this efficiently, the AQL query compiler interacts with the HDFS name node and attempts to co-locate query tasks with their associated input data.

\subsection{Transactions}\label{ss:tx}

As described in Section \ref{sec:dml}, AsterixDB supports record-level transactions across multiple LSM indexes in a Dataset. 
Concurrency control in AsterixDB is based on 2PL.
As transactions in AsterixDB just guarantee record-level consistency, all locks are node-local and no distributed locking is required.
Further, actual locks are only acquired for modifications of primary indexes and not for secondary indexes.  (Latches are employed to ensure atomicity of individual index operations.)
This allows for high concurrency on a variety of different index structures.
To avoid potential inconsistencies when reading the entries of a secondary index while the corresponding records of the primary index are being altered concurrently, secondary key lookups are always validated by (a) fetching the corresponding primary index entries while acquiring the necessary locks on the primary keys and (b) validating that the returned records are still consistent with the criteria for the secondary index search.
This post validation check can be seen in Figure \ref{fig:job}, where a select Operator is introduced to filter out any ``inconsistent'' records that are obtained from the primary index.

Recovery in AsterixDB employs a novel but simple method based on LSM-index-level logical logging and LSM-index component shadowing. 
For logical logging, the no-steal/no-force buffer management policy and write-ahead-log (WAL) protocols are followed, so each LSM-index-level update operation generates a single log record. 
For shadowing, when a new disk component is created by flushing an in-memory component or merging existing disk components, the new component is atomically installed by putting a validity bit into the component once the flush or merge operation has finished. 
Thus, only the committed operations from in-memory components need to be (selectively) replayed; any disk component without a validity bit is removed during crash recovery. More details on transactions in AsterixDB can be found in \cite{AsterixLSM}.

\subsection{Data Feeds}

AsterixDB is unique in providing ``out of the box'' support for continuous data ingestion.
Data Feed processing is initiated by the evaluation of a ^connect feed^ statement (see Section \ref{ss:feeds}).
The AQL compiler first retrieves the definitions of the involved components (feed, adaptor, function, and target Dataset) from the AsterixDB Metadata. 
It then translates the statement into a Hyracks Job that is scheduled to run on an AsterixDB cluster.
The dataflow described by this job is called a feed \emph{Ingestion Pipeline}. 
Like all Hyracks Jobs, an Ingestion Pipeline consists of Operators and Connectors.

To support cascading networks of feeds, an Ingestion Pipeline can provide \emph{Feed Joints} between Operators and Connectors. 
A Feed Joint is like a network tap and provides access to the data flowing along a pipeline. 
It adds a buffering capability for an Operator's output and offers a subscription mechanism and allows data to be routed simultaneously along multiple paths, e.g., to feed the Ingestion Pipeline of another feed.

A feed Ingestion Pipeline involves three Stages---\emph{intake}, \emph{compute} and \emph{store}. 
Each Stage corresponds to an Operator.
The \emph{Intake Operator} creates an instance of the associated feed adaptor, using it to initiate transfer of data and to subsequently transform the data into the ADM format. 
If the feed has an associated pre-processing function, it is applied to each feed record using an \emph{Assign Operator} as part of the compute Stage.
Finally, in the store Stage, the output records from the preceding intake/compute Stage are put into the target Dataset and secondary indexes (if any) using Hyracks' \emph{Insert Operators} for the individual index types.
More details, e.g. how dataflow is managed in large cascading networks of feeds or how fault scenarios are handled can be found in \cite{AsterixFeeds}.

\section{Status and performance}
\subsection{2013 and 2014 Releases}

There have been three public releases of AsterixDB to date.
The \emph{beta} release (0.8.0) appeared in June 2013.
It was the first release and showed -- not uncommon for first releases -- a lot of promise but some room for improvement.
Subsequent releases (0.8.3, 0.8.5) have come out at roughly five month intervals; these have been stabilization releases with improved performance plus a few minor features that our initial ``customers'' needed.

Two larger releases are planned for 2014.
Each will have one ``big'' feature as their main theme.
The first will make feeds available as described in this paper and in \cite{AsterixFeeds}.
The second release will add support for indexes over external Datasets, enabling
AsterixDB to index external data and use such indexes for query processing in the same way that is possible for internal Datasets today.

AsterixDB is in its childhood and has corresponding limitations.
One limitation, typical of early systems, is the absence of a cost-based query optimizer.
Instead, it has a set of fairly sophisticated but ``safe'' rules\footnote{The current AQL optimizer includes about 90 rules, with 50 coming from Algebricks and 40 being AQL-specific.} to determine the general shape of a physical query plan and its parallelization and data movement. 
The optimizer keeps track of data partitioning and only moves data as changes in parallelism or partitioning require.
Some examples of ``safe'' rewritings are (a) AsterixDB always chooses to use index-based access for selections if an index is available and (b) it always chooses parallel hash-joins over other join techniques for equijoins.
To give cost control to users, AsterixDB also supports query optimization hints.
Hints for
join methods,
grouping algorithms,
and overriding the unconditional use of index-based access paths are supported.

\begin{query}[Index hint]{q:hint}
for $user in dataset MugshotUsers
for $message in dataset MugshotMessages
where $message.author-id /*+ indexnl */ = $user.id
return {
  "uname" : $user.name,
  "message" : $message.message
};
\end{query}

Example Query \ref{q:hint} contains a hint suggesting that an index-based nested-loop technique should be used to process its join.

\subsection{Use Cases and Lessons}

Our ultimate goal in building AsterixDB was to build something scalable and ``cool'' that can solve Big Data related problems for actual users. 
To this end, we worked with some external users on pilot projects involving AsterixDB in mid/late 2013. 
We summarize them briefly here to share some things that were tried and the benefits that resulted for the AsterixDB effort from watching users work (and struggle) and then getting their feedback.

Our first pilot involved testing AsterixDB on a cell phone event analytics use case. 
This use case was similar in nature to click-stream analytics; it made heavy use of grouped aggregation to do windowed aggregation, came with a need to express predicates about the sequential order of certain events, and was envisioned as later being applied to very large Datasets. 
Learnings from this use case drove significant improvements in our parallelization of grouped aggregations and avoidance of materialization of groups that are formed only to be aggregated. 
It also drove us to add support for positional variables in AQL (akin to those in XQuery).

Another pilot, somewhat similar, involved social media (tweet) analytics. 
It was inspired by AsterixDB's capabilities around open Datatypes and grouped spatial aggregation and it involved being the back-end for an interactive analysis tool prototype. 
This pilot required dealing with a large volume of data and, being the first project to do that, exposed a series of issues in the AQL implementation. 
Most notably it unearthed issues related to materializing groups for aggregation, providing another driver for the materialization improvements in the second AsterixDB release.

A third pilot project involved using AsterixDB plus other statistical tools to do behavioral data analysis of information streams about events and stresses in the day of a 20-something computer and social media user. 
The data came from volunteer subjects' computer logs, heart rate monitors, daily survey responses, and entrance and exit interviews. 
This use case was ``small data'' for now, but drove improvements in our external data import capabilities and new requirements related to data output formats.
It further led us to add support for temporal binning, as time-windowed aggregation was needed. 
Also, one code refresh that we gave this group broke their system at an inopportune time (right before a paper deadline). 
An AsterixDB student had changed the system's metadata format in a way that made existing metadata incompatible with his change's expectations. 
This led to the user having to reload all of their data; it reminded us that we should eat our own dogfood (open types!) more heavily when managing our metadata.
\subsection{Current Performance}\label{sec:perf}

We now present some initial results on the performance of AsterixDB versus several widely used Big Data management technologies (one of each kind). 
A more comprehensive performance study is work in progress \cite{BDMS2}.

\subsubsection{Experimental Setup}\label{sec:perfsetup}

We ran the reported experiments on a 10-node IBM x3650 cluster with a Gigabit Ethernet switch. Each node had one Intel Xeon processor E5520 2.26GHz with four cores, 12GB of RAM, and four 300GB, 10K RPM hard disks. 
On each machine 3 disks were used for data (yielding 30 partitions). 
The other disk was used to store ``system data'' (e.g., transaction logs and system logs) if a separate log location is supported by the system.

The other systems considered are MongoDB\cite{mongoDB} (2.4.9, 64-bit), Apache Hive\cite{hive:website} (0.11 on Hadoop 1.0.4 using ORC files), and System-X, a commercial, shared-nothing parallel relational DBMS that others have also tested as ``System-X''.
To drive our experiments we used a client machine connected to the same Ethernet switch as the cluster nodes. For AsterixDB, we used its REST API to run queries. For System-X and Hive, their JDBC clients were used, and for MongoDB, we used its Java Driver (version 2.11.3).

We present results for a set of read-only queries and insert operations. 
We selected these queries and inserts to test a set of operations that most systems examined here support and thus provide a relative comparison of AsterixDB to each.
For the systems that support indexing, we use their version of B(\textsuperscript{+})-trees, as the test predicates do not involve advanced data types.
Note that the purpose of these tests is not to outperform other systems, but rather to show that AsterixDB does not sacrifice core performance while delivering its broader, ``one size fits a bunch'' feature set.
We should also note that there are a few cases where a given system does not support a query in a natural way:
Hive has no direct support for indexes, so it needs to scan all records in a table in cases where other systems use indexes;
in such cases we re-cite its unindexed query time.
MongoDB does not support joins, so we implemented a client-side join in Java to compare it to the other systems' join queries.
We report average response times on the client side as our performance metric. 
The reported numbers are based on running each workload 20 times, discarding the first five runs in calculating averages (considering them to be warm-up runs).

We conducted our experiments on a schema similar to the Section \ref{sec:dml} examples. 
Specifically, we used three datasets: users, messages and tweets, all populated with synthetic data. 
Table \ref{tab:sizes} shows the storage sizes. 
For AsterixDB, we report numbers for 2 different open data types: one that contains all the fields that are in the data (\emph{Schema}) and one that contains only the required fields (\emph{KeyOnly}, see Section \ref{ssec:dataverse}). 
We see the expected differences in dataset sizes in Table \ref{tab:sizes}.
For Hive, we used the ORC file format, which is a highly efficient way to store data. 
Hive benefits from the  compression of the ORC format at the storage level.
The test schema was very similar to the schema in Data definition \ref{ddl:types}, with the records having nested types. 
In AsterixDB and MongoDB we stored the records with nesting; 
we normalized the schema for System-X and Hive for the nested portions of the records.  
Our read-only queries were very similar to Section \ref{sec:dml}'s examples.
More details on the AsterixDB DDL and DML for all tests and on the client-side join implementation for MongoDB are available online at \cite{experiments}.

\subsubsection{Preliminary Results}\label{sec:perfresults}

Table \ref{tab:readOnly} presents the query response times from the various systems.
Below, we consider the queries and results in more detail.

\begin{table}[t]
\centering
\small
\begin{tabular}{|l||r|r|r|}
\hline
 & Users & Messages & Tweets  \\
\hline
Asterix (\emph{Schema})  &  192 & 120 & 330 \\
\hline
Asterix (\emph{KeyOnly}) & 360 & 240 & 600 \\
\hline
Syst-X & 290 & 100 & 495 \\
\hline
Hive & 38 & 12 & 25 \\
\hline
Mongo & 240 & 215 & 478 \\ 
\hline
\end{tabular}
\caption{Dataset sizes (in GB)}
\label{tab:sizes}
\vspace{-0.2in}
\end{table}

\begin{table}[t]
\centering
\small
\begin{tabular}{|l||r|r|r|r|r|}
\hline
 &\shortstack{Asterix\\\emph{Schema}}&\shortstack{Asterix\\\emph{KeyOnly}}&Syst-X&Hive\hspace*{1em}&Mongo\\
\hline
\hline
Rec Lookup     &   0.03 &   0.03 &   0.12 &   (379.11) &   0.02 \\
\hline
\hline
Range Scan     &  79.47 & 148.15 & 148.33 &  11717.18  & 175.84 \\
--- with IX    &   0.10 &   0.10 &   4.90 & (11717.18) &   0.05 \\
\hline
\hline
Sel-Join (Sm)  &  78.03 &  96.76 &  55.01 &    333.56  &  66.46 \\
--- with IX    &   0.51 &   0.55 &   2.13 &   (333.56) &   0.62 \\
\hline
Sel-Join (Lg)  &  79.62 &  99.73 &  56.65 &    350.92  & 273.52 \\
--- with IX    &   2.24 &   2.32 &  10.59 &   (350.92) &  14.97 \\
\hline
\hline
Sel2-Join (Sm) &  79.06 &  97.82 &  55.81 &    340.02  &  66.45 \\
--- with IX    &   0.50 &   0.52 &   2.62 &   (340.02) &   0.61 \\
\hline
Sel2-Join (Lg) &  80.18 & 101.24 &  56.10 &    394.11  & 313.17 \\
--- with IX    &   2.32 &   2.32 &  10.70 &   (394.11) &  15.28 \\
\hline
\hline
Agg (Sm)       & 128.66 & 232.30 & 130.64 &     83.18  & 400.97 \\
--- with IX    &   0.16 &   0.17 &   0.14 &    (83.18) &   0.19 \\
\hline
Agg (Lg)       & 128.71 & 232.41 & 132.19 &     94.11  & 401    \\
--- with IX    &   5.53 &   5.55 &   4.67 &    (94.11) &   8.34 \\
\hline
\hline
Grp-Aggr (Sm)  & 130.20 & 232.77 & 131.18 &    127.85  & 398.27 \\
--- with IX    &   0.45 &   0.46 &   0.17 &   (127.85) &   0.20 \\
\hline
Grp-Aggr (Lg)  & 130.62 & 234.10 & 133.02 &    140.21  & 400.10 \\
--- with IX    &   5.96 &   5.91 &   4.72 &   (140.21) &   9.03 \\
\hline
\end{tabular}
\caption{Average query response time (in sec)}
\label{tab:readOnly}
\vspace{-0.1in}
\end{table}

\begin{table}[t]
\centering
\small
\begin{tabular}{|c||c|c|c|c|}
\hline
 Batch Size & \shortstack{Asterix\\\emph{Schema}} & \shortstack{Asterix\\\emph{KeyOnly}} & Syst-X & Mongo \\
\hline
 1 & 0.091 & 0.093 & 0.040 & 0.035 \\
\hline
20 & 0.010 & 0.011 & 0.026 & 0.024 \\
\hline
\end{tabular}
\caption{Average insert time per record (in sec)}
\label{tab:insert}
\vspace{-0.2in}
\end{table}

The \emph{record lookup} query is a single-record selection that retrieves one complete record based on its primary key (similar to Query \ref{q:range-scan} but with a primary key predicate). 
System-X and Hive need to access more than one table to get all fields of the record since it has nested fields.
Hive also has to scan all records, which makes it the slowest by far for this query;
we put its response time in parentheses, as we wanted to show the number even though Hive is not designed for such queries.
Both AsterixDB and MongoDB utilize their primary indexes and can fetch all the fields of the record, including nested ones, without extra operations. 

The \emph{range scan} query is similar to Query \ref{q:range-scan}. 
It retrieves a small set of records with nested fields, using a range predicate on a temporal attribute to select the records. 
For System-X and Hive, small joins were needed to get the nested fields. 
We ran this query two ways:
once without a secondary index on the predicate attribute (forcing all systems to access all records), and once with a secondary index.
AsterixDB's response, using \emph{Schema} types, was faster than that of System-X (which needed to join) and MongoDB (which had no schema information).
With \emph{KeyOnly} types AsterixDB had to read more data and its response time was similar to System-X and MongoDB.
Exploiting a secondary index reduced the cost of query execution considerably in all systems, of course, as instead of all records, relatively few pages of data were accessed.

For the first join query, we picked a \emph{simple select join} query (similar to Query \ref{q:equijoin}) using a foreign-key relationship. 
We considered two cases for this query: with and without an index to support the foreign key lookups. 
Moreover, we ran this query in two versions with respect to the selectivity of the temporal selection predicate. 
In the large selectivity version, 3000 records passed the filter, while only 300 passed in the small selectivity version.
As mentioned earlier, for MongoDB we performed the join on the client side. 
Our client code finds the list of objectIds of matching documents based on the selection predicate and then performs a bulk lookup of this list on the other collection.
As Hive has no support for secondary indexes, we ran Hive only for the no-index case.
AsterixDB and System-X both used hybrid hash joins for the no-index case. 
In that case, AsterixDB's response time increases with \emph{KeyOnly} types due to the increased dataset sizes.
For the case with an index, the cost-based optimizer of System-X picked an index nested-loop join, as it is faster than a hash join in this case. 
In AsterixDB, our rule-based optimizer lets users provide a hint (see Query \ref{q:hint}) to specify the desired join method, so we used an index nested-loop join here as well.
For MongoDB, the client-side join is fine for a small number of documents, but it slows down significantly for a larger number of documents.
This appears to be due to both the larger number of documents it needs to access from both sides of the join and to the increased effort on the client side to compute the join results.

We also included a second join query, a \emph{double select join}, that has filter predicates on both sides of the join. 
Again, for this query, the table gives numbers for both selectivity versions, both without and with an index. 
The main difference between this query and the previous one is the second filter predicate, which reduces the final result size significantly.

The simple \emph{aggregation} query
in Table \ref{tab:readOnly}
is similar to Query \ref{q:agg}. 
It calculates the average message length in a given time interval. 
Again, two cases were considered: without and with an index supporting the filter on the dataset. 
We considered two versions of the query based on its filter selectivity (300 vs. 30000 records). 
For MongoDB, we needed to use its map-reduce operation for this query, as it could not be expressed using MongoDB's aggregation framework. 
Hive can only run the no-index case, but benefits from its efficient ORC file format which allows for a short data scan time. 
The bigger data size introduced by \emph{KeyOnly} data types in AsterixDB shows its impact again if no index is available to avoid a full dataset scan.
Using the index eliminated scans and improved response times for all systems that support indexing.
The response times for all systems that used an index are close to one another. 

We also considered a \emph{grouped aggregation} query similar to Query \ref{q:sort}. 
This query groups a subset of records in a Dataset, and within each group, uses an aggregate to rank and report the top 10 groups. 
Again, we ran this query without and with an index to select records, and with both small and large selectivities.
The group-by step and the final ranking necessitate
additional sorting and limiting of the result, when compared with the
simple aggregation query. As a result, we see a response
time increase for all systems.
For the indexed with small selectivity version, this increase is quite noticable in AsterixDB.
This appears to have two causes: 1) AsterixDB does not push limits into sort operations yet, and 2) the way AsterixDB fetches final partitioned query results introduces some overhead.
We are currently working on both issues.

Data ingestion is another important operation for a Big Data management system. 
Table \ref{tab:insert} shows the performance of the insert operation in the different systems. 
(Hive is absent because the life cycle for Hive data is managed outside the system.)
For MongoDB, we set the ``write concern'' to \emph{journaled} to provide the same durability as in AsterixDB and System-X. 
For a single record insert (batch size of 1), the current version of AsterixDB performs noticeably worse than MongoDB and System-X. 
This is mainly due to Hyracks job generation and start-up overheads.
By increasing the number of records inserted as a (one statement) batch, we can distribute this overhead to multiple records. 
With a batch size of 20, the average insert time per record in AsterixDB drops to a number that outperforms the other systems. 
Additionally, for use cases where an application needs to insert huge volumes of (possibly fast) data, AsterixDB provides bulk load and data feed capabilities that can offer even better performance. 
(The performance of feeds for providing continuous data ingestion is explored in \cite{AsterixFeeds}.)

To summarize these experimental findings, the results show that early AsterixDB is surprisingly competitive with more mature systems for various types of queries as well as for batched insertions. 
For queries that access all records (no indexing), AsterixDB’s times are close to
those of the other systems.
Similarly, when secondary indexes are used, AsterixDB enjoys the same benefits as the other systems that have indexing and therefore operates in the same performance ballpark.
Earlier sections of this paper showed the breadth of AsterixDB's features; these results show that providing this breadth does not require sacrificing basic system performance.
Based on these preliminary results, then, it appears that it may indeed be the case that ``one size fits a bunch''.
\section{Summary and future work}

This paper has provided the first complete and post-release description of AsterixDB, a new open source Big Data platform.
We have looked broadly at the system, covering its user model and feature set as well as its architectural components.
Unlike current Big Data platforms, AsterixDB is a full-function BDMS (emphasis on \emph{M}) that is best characterized as a cross between a Big Data analytics platform, a parallel RDBMS, and a NoSQL store, yet it is different from each.
Unlike Big Data analytics platforms, AsterixDB offers native data storage and indexing as well as querying of datasets in HDFS or local files; this enables efficiency for smaller as well as large queries.
Unlike parallel RDBMSs, AsterixDB has an open data model that handles complex nested data as well as flat data and use cases ranging from ``schema first'' to ``schema never''.
Unlike NoSQL stores, AsterixDB has a full query language that supports declarative querying over multiple data sets.

In addition to the above, AsterixDB features include a scalable new runtime engine; all-LSM-based data storage, with B+ tree, R tree, and keyword and n-gram indexes; a rich set of primitive types, including spatial, temporal, and textual types, to handle Web and social media data; support for fuzzy selections and joins; a built-in notion of data feeds for continuous ingestion; and NoSQL style ACIDity.
In initial tests comparing AsterixDB's performance to that of Hive, a commercial parallel DBMS, and MongoDB, 
AsterixDB fared surprisingly well for a new system. We are now working to improve and extend AsterixDB based on lessons stemming from that performance study as well as from pilot engagements with early users.  
Planned future work includes seamless integration with our Pregelix open source graph analytics system, potential use of HDFS for replicated LSM storage, and support for continuous queries and notifications to enable ``declarative pub/sub'' over Big Data.
AsterixDB is available for download at \cite{homepage}.
\vspace{0.1in}

\vspace{-1mm}
\noindent\textbf{Acknowledgments}
The AsterixDB project has been supported by a UC Discovery grant, NSF IIS awards 0910989, 0910859, 0910820, and 0844574, and NSF CNS awards 1305430, 1059436, and 1305253. 
The project has enjoyed industrial support from Amazon, eBay, Facebook, Google, HTC, Microsoft, Oracle Labs, and Yahoo!.
\vspace{-1mm}

\nocite{Onion12}
\nocite{Cattell11}
\nocite{mapred04}
\nocite{Cattell11}
\nocite{DeWitt:92}
\nocite{Economist-Big-Data}
\nocite{hive:website}
\nocite{hadoop:website}
\nocite{Gates09}
\nocite{lsm}
\nocite{calda}
\nocite{pavlo}
\nocite{NoOneSize}
\nocite{rares}
\nocite{xquery}
\nocite{json}
\nocite{docs}
\nocite{pregelix}
\nocite{compiler13}
\nocite{next13}
\nocite{vision}
\nocite{yingyi13}
\nocite{hyracks}
\nocite{AsterixWarehouse}
\nocite{AsterixLSM}
\nocite{AsterixFeeds}
\nocite{codebase}
\nocite{hivesterix}
\nocite{homepage}

\bibliographystyle{abbrv}
\bibliography{references}

\begin{thebibliography}{10}

\bibitem{Economist-Big-Data}
{Data, Data Everywhere}.
\newblock {\em The Economist}, February 25, 2010.

\bibitem{AsterixLSM}
S.~Alsubaiee, A.~Behm, V.~Borkar, Z.~Heilbron, Y.-S. Kim, M.~Carey,
  M.~Dressler, and C.~Li.
\newblock {Storage Management in AsterixDB}.
\newblock {\em Proc. VLDB Endow.}, 7(10), June 2014.

\bibitem{AsterixWarehouse}
S.~Alsubaiee, A.~Behm, R.~Grover, R.~Vernica, V.~Borkar, M.~Carey, and C.~Li.
\newblock {ASTERIX: Scalable Warehouse-style Web Data Integration}.
\newblock {\em IIWeb}, pages 2:1--2:4, 2012.

\bibitem{vision}
A.~Behm, V.~Borkar, M.~Carey, R.~Grover, C.~Li, N.~Onose, R.~Vernica,
  A.~Deutsch, Y.~Papakonstantinou, and V.~Tsotras.
\newblock {ASTERIX: Towards a Scalable, Semistructured Data Platform for
  Evolving-world Models}.
\newblock {\em Distributed and Parallel Databases}, 29(3):185--216, 2011.

\bibitem{compiler13}
V.~Borkar and M.~Carey.
\newblock {A Common Compiler Framework for Big Data Languages: Motivation,
  Opportunities, and Benefits.}
\newblock {\em IEEE Data Eng. Bull.}, 36(1):56--64, 2013.

\bibitem{hyracks}
V.~Borkar, M.~Carey, R.~Grover, N.~Onose, and R.~Vernica.
\newblock {Hyracks: A Flexible and Extensible Foundation for Data-intensive
  Computing}.
\newblock {\em ICDE}, 0:1151--1162, 2011.

\bibitem{next13}
V.~Borkar, M.~Carey, and C.~Li.
\newblock {Big Data Platforms: What's Next?}
\newblock {\em XRDS}, 19(1):44--49, Sept. 2012.

\bibitem{Onion12}
V.~Borkar, M.~Carey, and C.~Li.
\newblock {Inside "Big Data Management": Ogres, Onions, or Parfaits?}
\newblock {\em EDBT}, pages 3--14, 2012.

\bibitem{DBLP:journals/corr/abs-1203-0160}
Y.~Bu, V.~Borkar, M.~Carey, J.~Rosen, N.~Polyzotis, T.~Condie, M.~Weimer, and
  R.~Ramakrishnan.
\newblock {Scaling Datalog for Machine Learning on Big Data}.
\newblock {\em CoRR}, abs/1203.0160, 2012.

\bibitem{yingyi13}
Y.~Bu, V.~Borkar, G.~Xu, and M.~Carey.
\newblock {A Bloat-aware Design for Big Data Applications}.
\newblock {\em ISMM}, 2013.

\bibitem{Cattell11}
R.~Cattell.
\newblock {Scalable SQL and NoSQL Data Stores}.
\newblock {\em SIGMOD Rec.}, 39(4):12--27, May 2011.

\bibitem{mapred04}
J.~Dean and S.~Ghemawat.
\newblock {MapReduce: Simplified Data Processing on Large Clusters}.
\newblock {\em OSDI}, pages 137--150, 2004.

\bibitem{DeWitt:92}
D.~DeWitt and J.~Gray.
\newblock {Parallel Database Systems: The Future of High Performance Database
  Systems}.
\newblock {\em Commun. ACM}, 35(6):85--98, June 1992.

\bibitem{Gates09}
A.~Gates, O.~Natkovich, S.~Chopra, P.~Kamath, S.~Narayanamurthy, C.~Olston,
  B.~Reed, S.~Srinivasan, and U.~Srivastava.
\newblock {Building a High-level Dataflow System on Top of Map-Reduce: The Pig
  Experience}.
\newblock {\em Proc. VLDB Endow.}, 2(2):1414--1425, Aug. 2009.

\bibitem{AsterixFeeds}
R.~Grover and M.~Carey.
\newblock {Scalable Fault-Tolerant Data Feeds in AsterixDB}.
\newblock {\em CoRR}, abs/1405-1705, 2014.

\bibitem{FMC}
F.~Keller and S.~Wendt.
\newblock {FMC: An Approach Towards Architecture-Centric System Development}.
\newblock In {\em ECBS}, pages 173--182, 2003.

\bibitem{Pregel}
G.~Malewicz, M.~H. Austern, A.~J.~C. Bik, J.~C. Dehnert, I.~Horn, N.~Leiser,
  and G.~Czajkowski.
\newblock {Pregel: a system for large-scale graph processing}.
\newblock In {\em SIGMOD Conference}, 2010.

\bibitem{lsm}
P.~O'Neil, E.~Cheng, D.~Gawlick, and E.~O'Neil.
\newblock {The Log-Structured Merge-Tree ({LSM}-tree)}.
\newblock {\em Acta Inf.}, 33:351--385, June 1996.

\bibitem{calda}
A.~Pavlo, E.~Paulson, A.~Rasin, D.~Abadi, D.~DeWitt, S.~Madden, and
  M.~Stonebraker.
\newblock {A Comparison of Approaches to Large-scale Data Analysis}.
\newblock In {\em SIGMOD}, pages 165--178, 2009.

\bibitem{BDMS2}
P.~Pirzadeh, T.~Westmann, and M.~Carey.
\newblock {A Performance Study of Big Data Management Systems}.
\newblock {\em in preparation}.

\bibitem{pavlo}
M.~Stonebraker, D.~Abadi, D.~DeWitt, S.~Madden, E.~Paulson, A.~Pavlo, and
  A.~Rasin.
\newblock {MapReduce} and parallel {DBMS}s: Friends or foes?
\newblock {\em Commun. ACM}, 53:64--71, 2010.

\bibitem{NoOneSize}
M.~Stonebraker and U.~Cetintemel.
\newblock {One Size Fits All: An Idea Whose Time Has Come and Gone}.
\newblock {\em ICDE}, 0:2--11, 2005.

\bibitem{rares}
R.~Vernica, M.~Carey, and C.~Li.
\newblock {Efficient Parallel Set-similarity Joins Using MapReduce}.
\newblock {\em SIGMOD}, pages 495--506, 2010.

\bibitem{codebase}
{AsterixDB Code Repository}.
\newblock \url{https://code.google.com/p/asterixdb}.

\bibitem{docs}
{AsterixDB Documentation}.
\newblock \url{http://asterixdb.ics.uci.edu/documentation/}.

\bibitem{experiments}
{Experiment Details}.
\newblock \url{https://asterixdb.ics.uci.edu/pub/asterix14/experiments.html}.

\bibitem{hadoop:website}
Apache {{Hadoop}}.
\newblock \url{http://hadoop.apache.org/}.

\bibitem{hive:website}
Apache {{Hive}}.
\newblock \url{http://hive.apache.org/}.

\bibitem{hivesterix}
{Hivesterix}.
\newblock \url{http://code.google.com/p/hyracks/wiki/HivesterixUserManual028}.

\bibitem{homepage}
{AsterixDB}.
\newblock \url{http://asterixdb.ics.uci.edu/}.

\bibitem{json}
{JSON}.
\newblock \url{http://www.json.org/}.

\bibitem{mongoDB}
{MongoDB}.
\newblock \url{http://www.mongodb.org/}.

\bibitem{pregelix}
{Pregelix}.
\newblock \url{http://hyracks.org/projects/pregelix/}.

\bibitem{VXQuery}
{Apache VXQuery Incubating}.
\newblock \url{http://incubator.apache.org/vxquery}.

\bibitem{xquery}
{XQuery} 1.0: An {XML} query language.
\newblock \url{http://www.w3.org/TR/xquery/}.

\end{thebibliography}


\end{document}